\documentclass[aps,prl,twocolumn,superscriptaddress,amsmath,showpacs]{revtex4-1}

\usepackage{graphicx}
\usepackage{amsmath}
\usepackage{amssymb}

\begin{document}

\title{Gaussian error correction of quantum states in a correlated noisy channel}



\author{Mikael Lassen*}
\affiliation{Department of Physics, Technical University of Denmark, Fysikvej, 2800 Kongens Lyngby, Denmark}

\author{Adriano Berni}
\affiliation{Department of Physics, Technical University of Denmark, Fysikvej, 2800 Kongens Lyngby, Denmark}

\author{Lars S. Madsen}
\affiliation{Department of Physics, Technical University of Denmark, Fysikvej, 2800 Kongens Lyngby, Denmark}

\author{Radim Filip}
\affiliation{Department of Optics, Palack\' y University, 17. listopadu 12,  771~46 Olomouc, Czech Republic}

\author{Ulrik L. Andersen}
\affiliation{Department of Physics, Technical University of Denmark, Fysikvej, 2800 Kongens Lyngby, Denmark}

\date{\today}


\begin{abstract}
Noise is the main obstacle for the realization of fault tolerant quantum information processing and secure communication over long distances.
In this work, we propose a communication protocol relying on simple linear optics that optimally protects quantum states from non-Markovian or correlated noise. We implement the protocol experimentally and demonstrate the near ideal protection of coherent and entangled states in an extremely noisy channel. Since all real-life channels are exhibiting pronounced non-Markovian behavior, the proposed protocol will have immediate implications in improving the performance of various quantum information protocols.
\end{abstract}

\maketitle

A future quantum information network will consist of quantum communication channels that connect different nodes of the network \cite{Kimble2008}. These quantum links could either be used for establishing a secret key between nodes, and thereby allowing for unconditional secure communication, or they could be used for communication of quantum information between quantum processors. The transmission of quantum information can be carried out either by sending the quantum states directly through the quantum links, or, by establishing entanglement between the nodes and subsequently use teleportation for transferring the quantum states~\cite{duan2005.nat}. The transmitted quantum information can be conveniently described by quantum states of two-level systems, that is, qubits, but a vast number of real world realizations rely on modes of the electro-magnetic fields described by quantum systems of continuous variables~\cite{Review1,Review2,Review3}. 

All these quantum communication schemes, however, will be ultimately limited in their performance by the noise that inevitably invades all realistic communication channels. Such noise may eventually lead to a lack of security in quantum key distribution (QKD) and to errors in directly transmitted quantum states. To combat the detrimental noise of the channel, various strategies have been proposed including noise-robust QKD protocols \cite{Patron2009,Pirandola2008,Madsen2012}, entanglement distillation protocols and error correcting codes \cite{shor1995.pra,steane1996.prl,laflamme1996.prl,bennett996.pra,cory1998.prl,Pittmann2005.pra,Chiaverini2004.nat,slotine1998.prl,braunstein1998.nat,duan2005.nat}. The complexity of these schemes strongly depends on the type of noise in the channel. It has been shown that if the noise is additive Gaussian,
and the information carrying states are Gaussian, then neither entanglement distillation nor quantum error correcting codes can be realized by simple Gaussian operations \cite{Eisert2002,Fiurasek2002,Giedke2002,Niset2009}.
On the other hand, for non-Gaussian error models (such as random attenuation and phase diffusion) simple Gaussian operations suffice to correct the errors in CV systems~\cite{braunstein1998.nat,Aoki2009,Niset2008,Lassen2010,Dong2008,Hage2008}. However, in many conventional communication systems, the error model is Gaussian, and thus it appears that one is faced with the complexity of implementing experimentally challenging non-Gaussian operations for enabling fault-tolerant quantum communication~\cite{Eisert,Takahashi2010,Ralph2011,Micuda2012}.

In the above mentioned No-Go theorems, the Gaussian noise is assumed to be uncorrelated.
However, with the miniaturization of solid state systems and the increasing speed of optical communication, the noise in todays communication systems inevitably exhibit correlations in time and space \cite{Kretschmann2005,Corney2006}, and thus it will be relevant to consider channels with correlated noise. In this case, the No-Go theorem does not apply.
Here we propose a simple encoding and decoding technique based on linear optical transformations that ideally protects {\em arbitrary} quantum states from Gaussian noise in correlated quantum channels. The protocol works in particular for Gaussian quantum states, and thus, perfect Gaussian error correction with Gaussian transformations is possible due to the correlations of the channel noise. We implement the protocol for coherent and entangled states of light, thereby characterizing the protocol for the two main communication approaches; direct communication and teleportation based communication. Correlated noise in quantum communication were initially considered for qubits leading to the concept of decoherence free subspace \cite{Lidar2003,Kwiat2000,Banaszek} but recently a few theoretical studies have also addressed correlated noise in bosonic channels using similar strategies~\cite{Giovannetti2005,Cerf2005,Lupo2010,Caruso2012}.

\begin{figure}[h]
\begin{center}
\includegraphics[width=3.0in]{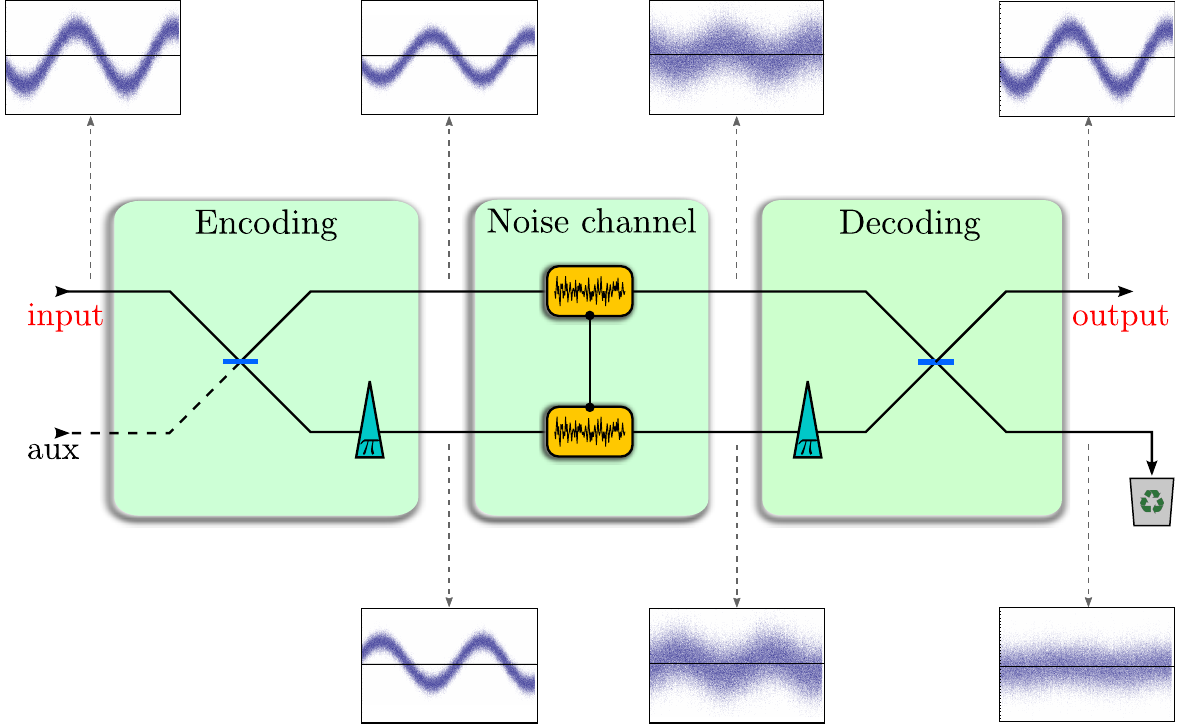}
\caption{Schematic of the proposed error correcting scheme for the protection of an arbitrary quantum state against correlated noise. The scheme is divided into three different steps associated with the encoding, noise addition and decoding. The insets serve as an illustration of the function of the protocol on a coherent state. They are ensemble measurements of the coherent state quadratures as a function of time at different positions of the protocol. The input state exhibits quantum noise limited fluctuations but after the noisy channel, the state evidently contains excess noise. This is then removed at the correction stage, where the noise is clearly separated from pure quantum state.}
\label{theosetup}
\end{center}
\end{figure}

Our error-protecting scheme is depicted in Fig.~\ref{theosetup} for the case of two partially correlated channels with classical noise which could either represent two spatially separated channels with spatial correlations or two consecutive uses of a single channel with temporal correlations (corresponding to a non-Markovian channel). The channels are noisy and a part of the noise is described by the perfectly correlated complex random variables $\varepsilon_1$ and $\varepsilon_2$~\cite{Holevo2001}. To describe an asymmetry of the correlated classical noise among the two channels, we assume $\varepsilon_1=\sqrt{g_1}v_C$ and $\varepsilon_2=\sqrt{g_2}v_C$, where the magnitude of the classical excess noise contributions are given by the factors $g_1>0$ and $g_2>0$, and $v_C$ is the complex random variable corresponding to the classical fluctuations from the environment. Assuming the transmission of the channels to be $\eta$, the Bogoliubov transformations for the channels are
\begin{eqnarray}
\hat{a}'_{1}=\sqrt{\eta} \hat{a}_{1}+\sqrt{1-\eta} \hat{v}_{1}+\sqrt{g_1}v_C\\
\hat{a}'_{2}=\sqrt{\eta} \hat{a}_{2}+\sqrt{1-\eta} \hat{v}_{2}+\sqrt{g_2}v_C
\end{eqnarray}
where $\hat{a}'_{1,2}$ and $\hat{a}_{1,2}$  are the annihilation field operators associated with the input and output modes, and $\hat{v}_{1,2}$ represent the uncorrelated thermal fluctuations (which for a zero-temperature channel is identical to the loss-induced vacuum fluctuations). 
For circumventing the noise, we follow the encoding and decoding strategy illustrated in Fig. \ref{theosetup}. The channel inputs are prepared by combining the input signal ($\hat{b}_{in}$) with an auxiliary vacuum state ($\hat{b}_{aux}$) on a beam splitter, and subsequently introducing a relative phase shift of $\pi$ between the two resulting states. The encoding transformation can be written as $(\hat{a}_{1},\hat{a}_{2})=(\sqrt{T_e}\hat{b}_{in}-\sqrt{1-T_e}\hat{b}_{aux},-\sqrt{1-T_e}\hat{b}_{in}-\sqrt{T_e}\hat{b}_{aux})$
where $T_e$ is the transmissivity of the encoding beam splitter.
The decoding transformation is the reverse of the encoding transformation, and thus represented by the transformations; $(\hat{b}_{out},\hat{b}'_{aux})=(\sqrt{T_d}\hat{a}'_{1}-\sqrt{1-T_d}\hat{a}'_{2},-\sqrt{1-T_d}\hat{a}'_{1}-\sqrt{T_d}\hat{a}'_{2})$  where $T_d$ is the transmissivity of the decoding beam splitter.
By choosing $T_e=T_d=g_2/(g_1+g_2)$, the input-output relation for the entire scheme is
\begin{eqnarray}
\hat{b}_{out}=\sqrt{\eta} \hat{b}_{in}+\sqrt{1-\eta}\left(\sqrt{\frac{g_2}{g_1+g_2}}\hat{v}_{1}
-\sqrt{\frac{g_1}{g_1+g_2}}\hat{v}_{2}\right),\nonumber\\
\end{eqnarray}
which corresponds to a purely lossy but noiseless channel for any values of $g_1$ and $g_2$ (for the zero-temperature channel).
The correlated classical noise of the environment has therefore been {\it completely} separated from the signal; the noise will leave one output of the beam splitter whereas the signal will leave the other output. Even if the two channels are partially uncorrelated, our scheme is perfectly removing the correlated part of the noise without amplifying the uncorrelated part. For a general treatment of the protocol, see Supplemental Material \cite{supp}.


\begin{figure}[h]
\begin{center}
\includegraphics[width=3.0in]{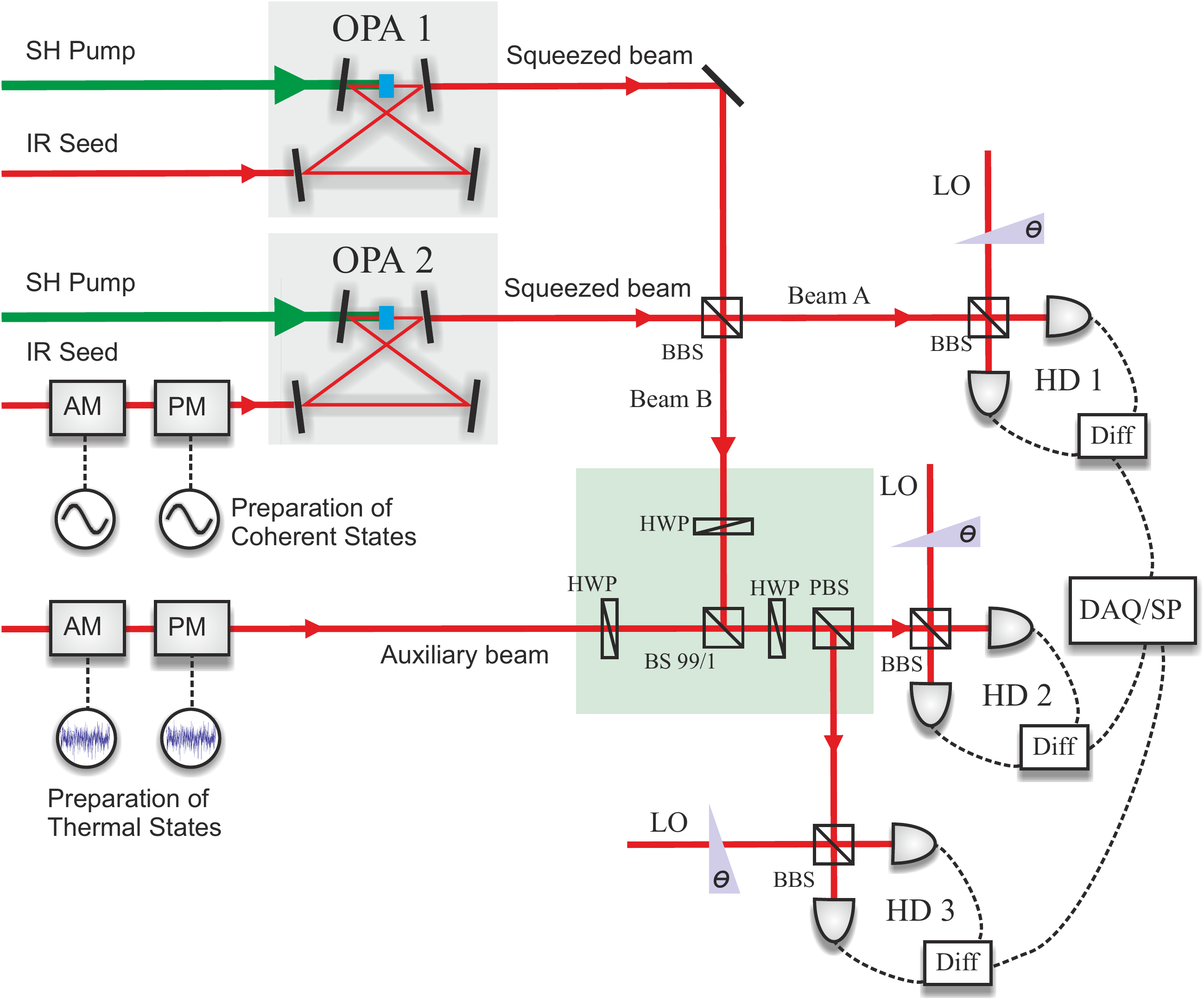}
\caption{Schematic of the experimental setup. The error correcting part is inside the shaded box which contains two input and two outputs: The inputs are for the input quantum state and for noise addition whereas the two outputs are associated with the two output of the protocol. The quantum states at the inputs are either coherent states or entangled states. Coherent states are prepared at the sideband frequencies of $\pm4.9$MHz using a pair of modulators; amplitude (AM) and phase modulators (PM), whereas the entangled states are generated by interfering two squeezed beams on a balanced beam splitter (BBS). The squeezed beams are produced by optical parametric amplifiers (OPA1 and OPA2) which are pumped by laser beams at 532nm (denoted "Pump") and stabilized with "seed" beams at 1064nm. When coherent states are used as inputs to the protocol, the OPAs are not operational and thus the coherent states produced in front of OPA2 are bypassing the squeezing operation and injected directly into the protocol. Correlated noise in the two channels is produced by an auxiliary beam that traverses a pair of noise-controlled modulators and subsequently injected into the scheme. For the verification, high-efficiency homodyne detectors (HD) are used; HD2 and HD3 are measuring the two outputs of the protocol whereas HD1 is used for the characterization of entanglement. PBS: Polarizing beam splitter; HWP: Half-wave plate; DAQ/SP: Data acquisition/signal processing. LO: Local oscillator.}
\label{expsetup}
\end{center}
\end{figure}

\begin{figure}[h]
\begin{center}
\includegraphics[width=3.0in]{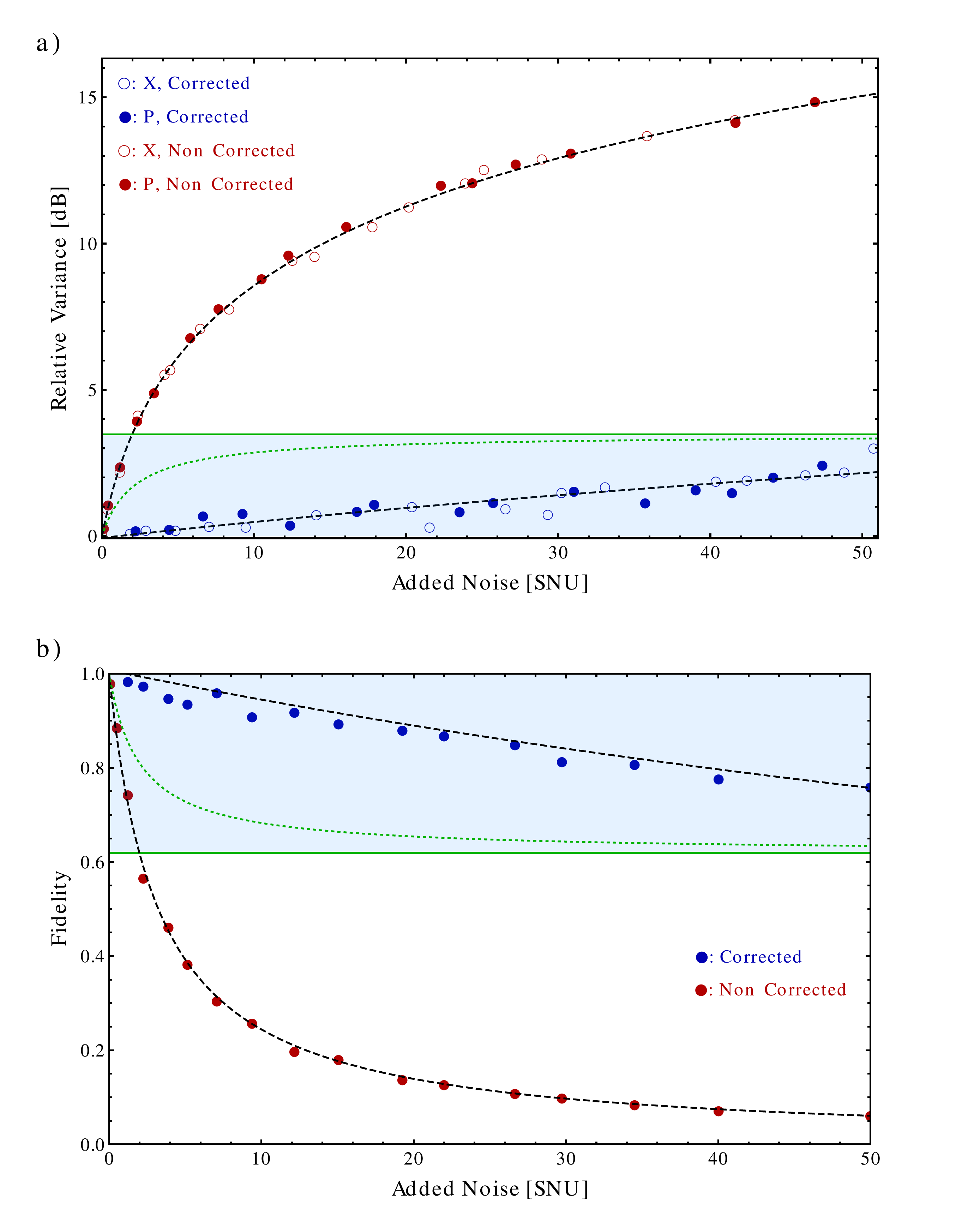}
\caption{Demonstration of error correction coding of a coherent state in an extremly noisy environment. (a) Relative noise variance (normalized to shot noise) of the quantum state is plotted against the channel excess noise for the case of slightly asymmetrically correlated noise with a ratio of $g_1/g_2=0.61$. Amplitude and phase quadratures are represented by open and closed circles, respectively, and the results before and after correction are shown. The non-corrected quantum states were measured by setting $T_d=1$ whereas the corrected states were measured with $T_d=0.36$. In (b) we plot the fidelity between the input coherent state and the output corrected and uncorrected states. The theoretical predictions (obtained using the theory outlined in \cite{supp}) are represented by the dashed lines, and the green (solid and dotted) lines correspond to the two different incoherent stategies depending on the apriori information about the channels. We use the solid line as our benchmark since in that case, the a priori channel information is similar to that needed for the implementing the coherent strategy \cite{supp}. The shaded parts therefore represent the regions at which the classical strategy is beaten. The small deviation from ideal performance at large added noise is due to the non-ideal mode matching at the correcting beam splitter. The theoretical line corresponds to a modemismatch of 1\% (corresponding to a visibility of 99.5\%). The statistical error bars are smaller than the dots.}
\label{EC_coh}
\end{center}
\end{figure}

It is important to note that the error protecting protocol is universal, i.e. it is valid for any input quantum state and for any statistics of the correlated noise. In the following we investigate our protocol experimentally for coherent and CV entangled states in a correlated Gaussian noisy environment, but the protocol would likewise work for two-dimensional qubit systems or non-Gaussian systems of higher dimensions.


The experimental realization of the scheme is illustrated in Fig. \ref{expsetup}.
The prepared quantum states - coherent and entangled states - are residing at the sideband frequencies of $\pm$4.9~MHz relative to the carrier frequency of the optical mode. Coherent states are produced by a pair of electro-optic modulators whereas the entangled states are generated by interfering squeezed beams on a balanced beam splitter.

We realize the two channels in two orthogonal polarization modes thereby simulating correlated polarization noise. In this basis, the input state is simply encoded by the use of a single half-wave plate which simulates a variable beam splitter and introduces a relative phase shift of $\pi$. Gaussian noise of the environment is generated by traversing a bright beam through a pair of electro-optical modulators that are driven by two independent electronic Gaussian noise sources. The noise is then subsequently fed into the two channels via an asymmetric beam splitter (99/1) at which it is coupled with the signal. The distribution of the noise among the two orthogonal polarization modes is carried out with a single half wave plate. All quantum states of the experiment are completely characterized using three pairs of homodyne detectors (see ref. \cite{supp} for details).

\begin{figure}[h]
\begin{center}
\includegraphics[width=3.0in]{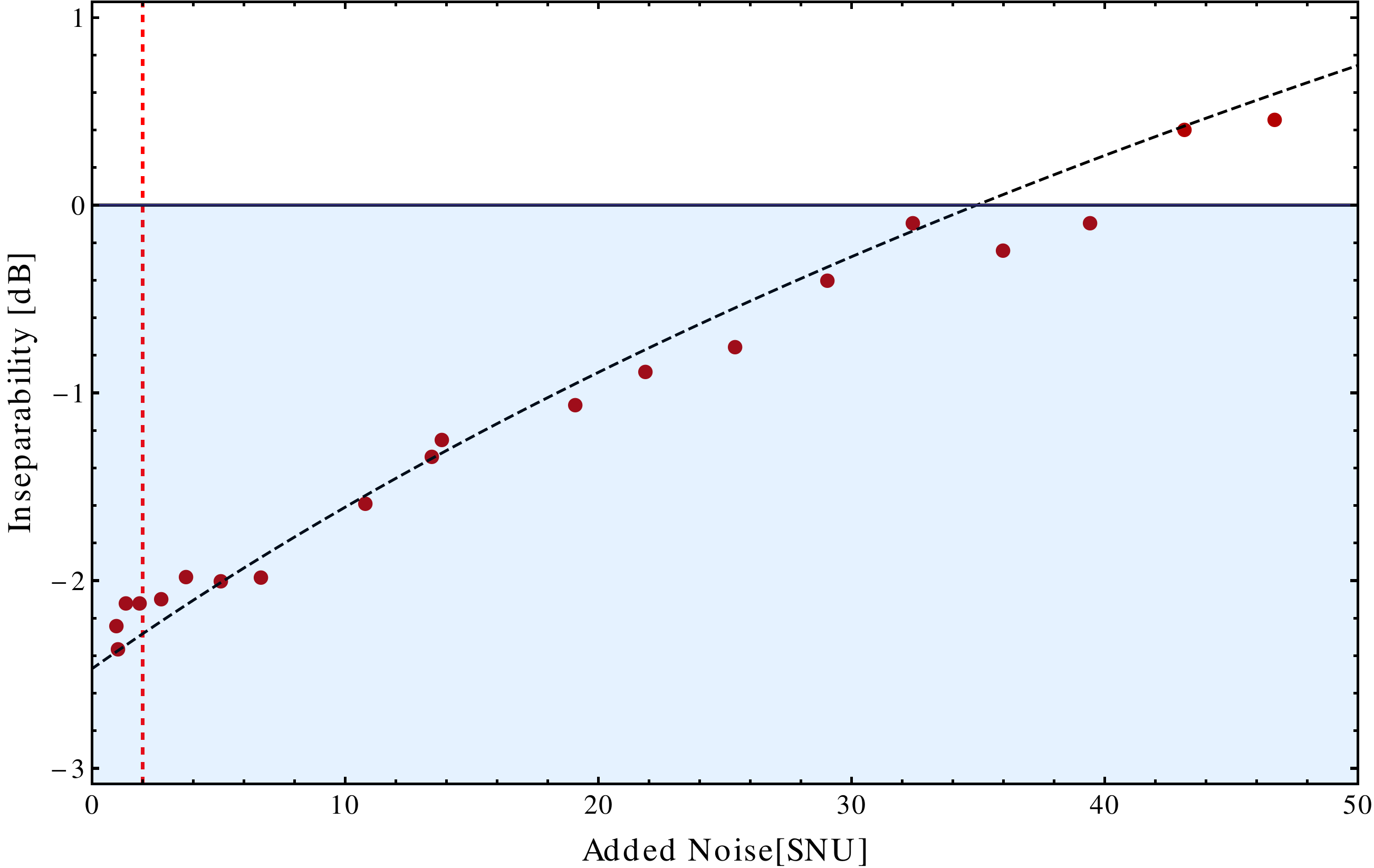}
\caption{Demonstration of error correction coding of continuous variable entangled states in an extremely noisy and symmetric channel. The inseparability number is plotted against the added channel noise relative to the shot noise limit (which is represented by the solid horizontal line). The vertical dashed line represent the entanglement breaking line associated with the point at which any input state will be unentangled if error correction is not used.
It is evident that the coherent strategy
keeps the state entangled (associated with the shaded region) for up to about 35 SNU of added noise, that is, far beyond the entanglement breaking point. The theory curve (dashed) takes into account the slight 1\% mode-mismatch at the decoding beam splitter, which then explains the increase of the two-mode squeezing variance as a function of the added noise. The statistical error bars are smaller than the dots.}
\label{EC_EPR}
\end{center}
\end{figure}

In Fig. \ref{theosetup}, we present the time traces recorded with a homodyne detector for the input coherent state, the noise-affected states of the channel, and the output states. It is evident from these traces that the noise of the channels is removed in the decoding station as a result of the coherent linear beam splitter interaction.
A quantitative study of the cancellation of excess noise is presented in Fig.~\ref{EC_coh} for coherent states. Here we plot the measured quadrature noise and fidelities between input and output states as a function of the channel noise with and without correction. Fig.~\ref{EC_coh}a represents the experimentally obtained variances for the realization corresponding to slightly asymmetric channels ($g_1/g_2=0.61$) with optimized encoding and decoding beam splitter settings; $T_e=T_d=0.38$. For this realization, the fidelities are displayed in Fig.~\ref{EC_coh}b. All data are compared to an incoherent (classical) correction protocol, where the quantum state is sent through one of the channels and the other channel (which contains correlated noise) is measured to correct the state. This incoherent strategy is independent on the amount of correlated noise as is the case for the coherent strategy \cite{supp}. We clearly see that the coherent approach beats the incoherent approach for all channel realizations.


\begin{figure}[h]
\begin{center}
\includegraphics[width=3.0in]{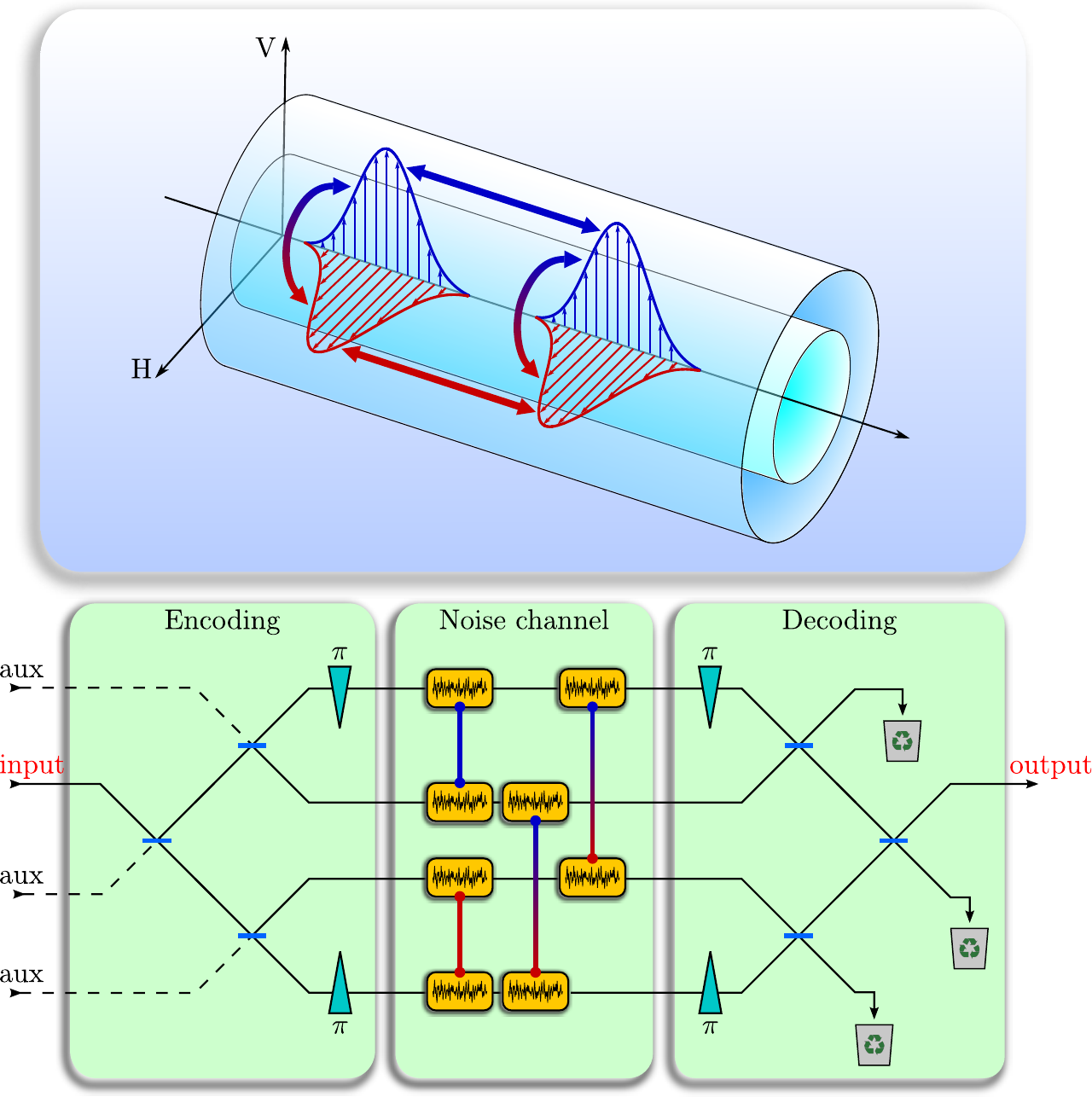}
\caption{The four-channel error correction coding scheme. a) Illustration of the pair-wise noise correlation pattern in an optical fiber. As a result of guided acoutic wave Brillouin scattering in the fiber, noise correlations (marked with arrows) arise in time as well as in the polarization during pulse propagation. b) Scheme for error correction of the correlated noise in fibers. The four pulses in a) are represented by four channels in b), and the noise correlations are marked by links between the channels and are color coded according to the colors in a).}
\label{4channel}
\end{center}
\end{figure}

Next we investigate the survival of entanglement in our correlated noisy channel. One half of a CV entangled state is sent through the noisy channel using the encoding-decoding protocol, and the resulting output measured with homodyne detection. The second half of the entangled state (which was not sent through the channel) is also measured and we compute the correlations in terms of variances of the joint quadratures; $\langle(\hat{x}_1-\hat{x}_2)^2\rangle$ and $\langle(\hat{p}_1+\hat{p}_2)^2\rangle$ where $\hat{x}$ and $\hat{p}$ represent the amplitude and phase quadrature of mode 1 and 2, obeying the relation $[\hat{x},\hat{p}]=i$. According to the criterion of Duan et al~\cite{Duan2000} and Simon~\cite{Simon2000}, entanglement is then present if $\langle(\hat{x}_1-\hat{x}_2)^2\rangle+\langle(\hat{p}_1+\hat{p}_2)^2\rangle<2$. The results of these measurements are displayed in Fig. \ref{EC_EPR}. The vertical line corresponds to the point at which a single channel no longer can be used for entanglement distribution. However, we clearly see that by the implementation of the error correcting code, the dual-channel can be used for deterministic entanglement distribution for at least up to 35 SNU of excess noise.



The residual excess noise after correction (evident in Fig. \ref{EC_coh} and \ref{EC_EPR}) stems from noise contributions that did not interfere at the decoding beam splitter as a result of the imperfect modematching at the noise-injecting beam splitter. Despite this small imperfection, we still show transmission of quantum states through an extremely noisy channel to a degree that allows for the generation of a secure key and the implementation of quantum teleportation. In addition to proving the channel's capability of transmitting entangled states, the results also indicate that the scheme is universal: Since a CV entangled state can be used to prepare an arbitrary state through state projection, the survival of entanglement unambigously proves the faithful transmission of a generalized state.

Protection of quantum states in correlated noisy environments is of practical relevance. One important example is the non-Markovian noise introduced by a standard optical fiber as a result of the effect of Guided Acoustic Wave Brilloiun Scattering (GAWBS)~\cite{Shelby1985}. The time scale of this noise is determined by the size of the fiber core, and in a standard fiber, the bandwidth is around a GHz. Therefore, for communication rates exceeding a GHz, consecutive pulses will contain correlated noise which can be cancelled using our protocol. It has also been shown that part of the GAWBS noise is depolarized, leading to correlations between orthogonal polarization modes as illustrated in Fig. \ref{4channel}. To simultaneously overcome both noise sources, we propose an extended version of our protocol as shown schematically in Fig. \ref{4channel}. The quantum state is divided into four channels: two channels are separated in time and two in polarization. Pairwise correlations between the four pulses occur due to polarized and depolarized GAWBS. A schematic of the separation and correlation is shown in Fig. \ref{4channel}b. In this realistic noise model, we protect the quantum states by adding $\pi$ phase shifts to two of the pulses with respect to the two others. After noise addition, the signal is then perfectly separated from the noise by following the linear interference strategy shown in Fig. \ref{4channel} \cite{supp}. If noise correlations exist in other degrees of freedom, it is possible to extend the protocol further. Our developed method for noise-protection therefore holds great promise for combating noise in real optical fiber systems as well as in miniaturized systems where spatial correlations exist.

In summary, we have proposed a universal scheme for protecting arbitrary quantum states in a noisy non-Markovian environment. The proposal has been investigated experimentally using coherent states and entangled states of light affected by correlated Gaussian noise.
Using a simple linear optical encoding and decoding scheme, we have demonstrated the near ideal recovery of pure quantum states from a highly noisy environment. The scheme can be easily extended to encompass four correlated noise channels which is of relevance for protecting quantum states in realistic optical fibers. This method has the potential to extend the distant for quantum communication and at which a secret key can be generated.

The work was financed by the Danish Agency for Science, Technology and Innovation (Sapere Aude projects no. 10-093584 and no. 10-081599). 
R.F. acknowledges financial support from grant P205/12/0577 of Czech Science Foundation.


\end{document}